\documentclass[aps,prb,amsmath,twocolumn]{revtex4}
\usepackage{bm}
\usepackage{graphicx}
\begin{document}
\title{Current Fluctuations and Electron-Electron Interactions in
  Coherent Conductors}

\author{Artem V. Galaktionov$^{1,3}$, Dmitri S. Golubev$^{2,3}$, and
Andrei D. Zaikin$^{1,3}$}
\affiliation{$^{1}$Forschungszentrum Karlsruhe, Institut f\"ur Nanotechnologie,
76021, Karlsruhe, Germany\\
$^2$Institut f\"ur Theoretische Festk\"orperphysik,
Universit\"at Karlsruhe, 76128 Karlsruhe, Germany \\
$^{3}$I.E. Tamm Department of Theoretical Physics, P.N.
Lebedev Physics Institute, 119991 Moscow, Russia}

\begin{abstract}
We analyze current fluctuations in mesoscopic coherent conductors
in the presence of electron-electron interactions.
In a wide range of parameters we obtain explicit
universal dependencies of the current noise on temperature,
voltage and frequency.
We demonstrate that Coulomb interaction decreases the
Nyquist noise. In this case the interaction correction
to the noise spectrum is governed by the combination
$\sum_nT_n(T_n-1)$, where $T_n$ is the transmission of the
$n$-th conducting mode. The effect
of electron-electron interactions on the shot noise is more
complicated. At sufficiently large voltages we recover two
different interaction corrections entering with opposite signs.
The net result is proportional to $\sum_nT_n(T_n-1)(1-2T_n)$,
i.e. Coulomb interaction decreases the shot noise at low transmissions
and increases it at high transmissions.

\end{abstract}
\maketitle

\section{Introduction}
Recent advances in nanotechnology enable
detailed investigations of a variety of quantum effects in mesoscopic
conductors. These investigations are of primary interest because
of fundamental importance of such effects as well due to rapidly
growing number of their potential applications. A great deal of
information is usually obtained from studying electron transport.
Additional/complementary information can be extracted
from investigations
of fluctuation effects, such as shot noise \cite{bb}. For instance,
it was demonstrated \cite{Khl,Les,Bu1} that the power spectrum of
the shot noise in coherent mesoscopic
conductors is expressed in terms of the parameter
\begin{equation}
\beta=\frac{\sum_n T_n(1-T_n)}{\sum_n T_n}. \label{landr}
\end{equation}
Here and below $T_n$ stands for the transmission of the $n$-th
conducting channel of a coherent conductor. Thus, since
transport measurements only allow to determine
the combination
\begin{equation}
\frac{1}{R}=\frac{2e^2}{h}\sum_nT_n,
\label{Landauer}
\end{equation}
studies of
the shot noise provide additional valuable information about the
transmission distribution of conducting modes.

The above results apply to the situations when interaction
between electrons can be neglected. In the presence of
electron-electron interactions the Landauer conductance
(\ref{Landauer}) and the $I-V$ curve are modified in a non-trivial
way. Recently it was shown \cite{GZ00} that the $I-V$ curve of a
(comparatively short) coherent conductor with arbitrary transmission
distribution $T_n$ in the presence
of interactions can be expressed in the form
\begin{equation}
R\frac{dI}{dV}=1-\beta f(V,T),
\label{univ}
\end{equation}
where $f(V,T)$ is a {\it universal} function to be defined below.
This result holds in the limit of large conductances $R \ll
R_q=h/e^2$ or, otherwise, at sufficiently high temperatures/voltages.
It demonstrates that the magnitude of the interaction correction
is controlled by {\it the same} parameter $\beta$ (\ref{landr})
which is already well known in the theory of shot noise. Physically
this result can easily be understood since both phenomena are related to
discrete nature of the electron charge. Hence,
there exists a direct link between shot noise and interaction
effects in mesoscopic conductors.

It is obvious that not only the $I-V$ curve (\ref{univ}) but also
shot noise as well as higher moments of the current operator should
be affected by electron-electron interactions. This paper is devoted
to a detailed investigation of current fluctuations in mesoscopic
coherent conductors in the presence of electron-electron
interactions.
Previously various aspects of this problem have been studied for a particular
case of tunnel junctions in the Coulomb blockade regime, see, e.g.
Refs. \onlinecite{BMS,Yuli,ZGP,LL}. The effect of
interactions on shot noise in 2d diffusive conductors at sufficiently
high temperatures was recently addressed in Ref. \onlinecite{Gefen}.

Here we will employ a model of a coherent con\-duc\-tor \cite{GZ00,Naz,GZ02}.
Within this model we will demonstrate that interactions lead to two different
corrections to the shot noise spectrum. One of these corrections scales with
the parameter $\beta$ (\ref{landr}). This correction is always {\it negative},
similarly to that found in Eq. (\ref{univ}) for the $I-V$ curve. It describes
(partial) suppression of the current noise due to Coulomb blockade. In
addition to this correction we shall find another one, which is proportional
to the parameter
\begin{equation}
\gamma=\frac{\sum_n T_n^2(1-T_n)}{\sum_n T_n} .\label{gam}
\end{equation}
This second correction is {\it positive}, i.e. it leads to relative
enhancement of the shot noise. The latter
correction turns out to be important only at voltages exceeding both
frequency and temperature and is negligible otherwise. Thus, at
sufficiently high voltages two interaction corrections -- negative
and positive -- compete, for $\gamma >\beta /2$ the second one wins
and, hence, in this case an overall enhancement of the shot noise by
interactions is predicted.

Our paper is organized as follows. In section 2 we will describe the model and
highlight our key results. A detailed derivation of these results is then
performed in section 3. Our main conclusions are briefly summarized in section
4. Most of the technical details, such as the derivation of our effective
action as well as a few other issues are presented in Appendices A,B and C.

\section{The Model and key results}

Similarly to Refs. \onlinecite{GZ00,Naz,GZ02} we will consider a
coherent scatterer between two big reservoirs. The scatterer is
described by an arbitrary distribution of transmissions $T_n$ of
its conducting modes, and the corresponding Landauer conductance
$1/R$ is defined in Eq. (\ref{Landauer}). The transmissions
$T_n$ are assumed to be energy independent. Phase and energy
relaxation may take place in these reservoirs but not inside the
scatterer, i.e. the scatterer is assumed to be shorter than both
dephasing and inelastic lengths. The scatterer region has an effective
capacitance $C$. For simplicity we will also assume that the charging
energy $E_C=e^2/2C$ does not exceed the typical inverse traversal
time. The scatterer is connected to the voltage source $V_x$ via linear
external impedance $Z_S$ (see Fig. 1). Here we restrict ourselves to a
simple case $Z_S(\omega )=R_S$. If necessary, generalization of our calculation
to arbitrary $Z_S(\omega )$ can be performed in a straightforward manner.

\begin{figure}
\includegraphics[width=3in]{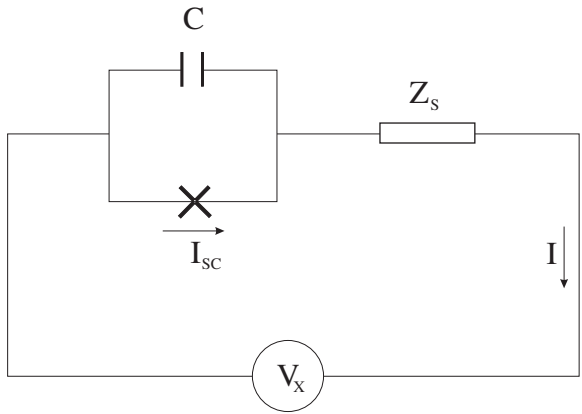}
\caption{The circuit under consideration. The scatterer (denoted by a
  cross) has
a capacitance $C$ and is connected to the voltage source $V_x$ via an
impedance $Z_S$.}
\end{figure}

In what follows we will investigate the current noise and evaluate
the correlation function
\begin{equation}
 {\cal S}(t,t')=\frac{1}{2}\langle \hat I(t)\hat I(t')+\hat I(t')\hat I(t)
 \rangle -\langle \hat I \rangle^2,
\label{corr}
\end{equation}
where $\hat I$ is the current operator in the circuit of Fig. 1. This
correlator can be expressed in the form
\begin{equation}
{\cal S}(t,t')={\cal S}^{\rm ni}(t,t')+\delta {\cal S}(t,t'),
\label{int}
\end{equation}
where ${\cal S}^{\rm ni}$ is the noninteracting contribution to
the current noise
\cite{Khl,Les,Bu1} and $\delta {\cal S}$ is the correction due to
electron-electron interactions inside the scatterer. This correction
will be evaluated in the most interesting ``metallic'' limit
\begin{equation}
g_0=g+g_S \gg 1.
\label{metal}
\end{equation}
Here we introduced dimensionless conductances of the scatterer and
the shunt, respectively $g=R_q/R$ and $g_S=R_q/R_S$. Eq. (\ref{metal})
implies that at least one of these two dimensionless conductances
is required to be much larger than unity.

Quite obviously, the latter correlator (\ref{corr}) should depend on
both $R$ and $R_S$.
We also introduce another correlator $\tilde
{\cal S}(t,t')$
defined by the same Eq. (\ref{corr}) in which one should substitute
the current operator across the scatterer $\hat I \to \hat I_{\rm
  sc}$. The two
correlators  ${\cal S}(t,t')$ and $\tilde {\cal S}(t,t')$ are not independent.
With the aid of the current conservation condition and performing
the Fourier transformation with respect to $t-t'$, one easily finds
the relation
\begin{eqnarray}
\tilde {\cal S}_\omega &=&\frac{R_S^2}{R_0^2}
(1+\omega^2R_0^2C^2){\cal S}_\omega
\nonumber\\
&& -\,\frac{R_S}{R^2}(1+\omega^2R^2C^2)\omega
\coth\frac{\omega}{2T},
\label{rel1}
\end{eqnarray}
where $R_0=RR_S/(R+R_S).$ The second term  is due
the noise produced by the external resistor $R_S$ which has to be
subtracted in order to arrive at $\tilde {\cal S}_\omega$.

In general also the correlator $\tilde {\cal S}_\omega$ depends on both $R$
and $R_S$. However, in the limit $R_S \gg R$ the dependence on the shunt
resistance is weak and can be neglected. In this case the interaction
correction to the current noise spectrum $\delta \tilde {\cal S}_\omega$
depends only on the properties of the scatterer. Below we will
present our key results for  $\delta \tilde {\cal S}_\omega$ only in this
limit. More general expressions can be found in Sec. 3.

Let us define the average voltage across the scatterer
$V=V_xR/(R+R_S)$ and consider first the limit of relatively small voltages.
At sufficiently large temperatures and/or frequencies we find
\begin{eqnarray}
\delta \tilde {\cal S}_\omega=-\frac{2\beta E_C}{3R},&&\text{if}\;
 T\gg gE_C,|eV|,|\omega| ,
\label{highT} \\
\delta \tilde {\cal S}_\omega=-\frac{\beta E_C }{R},&&\text{if}\;
|\omega|\gg T,gE_C,|eV|.
\end{eqnarray}

At lower temperatures and frequencies we obtain
\begin{eqnarray}
\delta \tilde {\cal S}_\omega=-\frac{4\beta T}{R_q}\ln\frac{gE_C}{T},\quad
\text{if}\; |\omega|,|eV|\ll T\ll gE_C,\label{intT}
\\ \delta \tilde {\cal S}_\omega=-\frac{2\beta |\omega|}{R_q}\ln
\frac{gE_C}{|\omega|}, \quad\text{if}\; T,|eV|\ll |\omega|\ll gE_C .
\end{eqnarray}
These results apply as long as either temperature or frequency exceeds
an exponentially small parameter $gE_C\exp (-g/2 )$. For even smaller
frequencies and temperatures \cite{FN} we get
\begin{equation}
\delta \tilde {\cal S}_\omega=-\frac{\beta \omega}{R}\coth \frac{\omega}{2T} .
\label{lowT}
\end{equation}
Note that the above expressions could also be anticipated from the
fluctuation-dissipation theorem (FDT) combined with the results
\cite{GZ00}. Indeed, in the limit of low voltages the current
noise is described by the standard Nyquist formula. Hence, in order to
satisfy FDT one should simply substitute the effective conductance
(\ref{univ}) into this formula. In this way one gets the interaction
correction $\delta \tilde S_\omega$ proportional to $\beta f$. For
instance, in the low frequency limit one finds \cite{GZ00}
$f(0,T)\simeq E_C/3T$ for $T \gg gE_C$, $f(0,T) \simeq (2/g)\ln
(gE_C/T)$ for $\exp (-g/2) \ll T/gE_C \ll 1$ and $f(0,T) \simeq 1$ for
$T < gE_C\exp (-g/2)$. Combining these expressions with FDT one
immediately reproduces Eqs. (\ref{highT}), (\ref{intT}) and
(\ref{lowT}).

It is worth stressing that here we evaluate
the current-current correlation functions directly and {\it do not} use
the results \cite{GZ00} together with FDT. However, it is satisfactory
to observe that FDT is explicitly maintained in our calculation
and the results derived here are fully consistent with those of Ref.
\onlinecite{GZ00}.

Now let us turn to the case of relatively large voltages $V$ where the
shot noise becomes important. As it was already announced, in this
case the correction to the noise power spectrum is proportional
to the parameter
\begin{equation}
\beta-2\gamma =\frac{\sum_nT_n(1-T_n)(1-2T_n)}{\sum_nT_n}.
\label{bg}
\end{equation}
In particular we obtain
\begin{eqnarray}
\delta \tilde {\cal S}_\omega=-\frac{2(\beta-2\gamma)|eV|}{R_q}\ln
\frac{gE_C}{|eV|}, \\ \text{if}\; T,|\omega|\ll |eV|\ll gE_C, \nonumber\\
\delta \tilde S_\omega=-\frac{(\beta-2\gamma)E_C}{R},\quad \text{if}\;|eV|\gg
T,gE_C,|\omega|.
\end{eqnarray}
We note that this correction can be either negative or positive
depending on the relation between the parameters $\beta$ and $\gamma$.
Thus, in contrast to the limit of low voltages (Nyquist noise),
one {\it cannot} conclude
that shot noise is always reduced by interactions. This reduction
occurs only for conductors with relatively low transmissions $\beta >
2\gamma$, while for systems with higher transmissions the net effect
of the electron-electron interaction enhances the shot noise.
In the important case of diffusive conductors one has
$\beta =1/3$, $\gamma =2/15$ and, hence,
$$
\beta -2\gamma =\frac{1}{15}.
$$
In this case the shot noise is reduced by interactions.

The above results have a transparent physical interpretation. At low
voltages the power spectrum of the Nyquist noise is proportional to
the system conductance $\propto \sum_nT_n$. Since in the presence
of interactions the conductance acquires a correction proportional to
$\beta$, the interaction correction to the Nyquist noise should
scale with the same parameter (\ref{landr}). On the other hand, shot
noise is determined by the combination
$\sum_n(T_n-T_n^2)$. Accordingly, the interaction correction to the
shot noise power should consist of two contributions. One of
them comes from $\sum_nT_n$ and is again proportional to $\beta$.
Another contribution originates from the interaction correction
to  $\sum_nT_n^2$ which turns out to scale as $2\gamma$.
Since these two corrections enter with the opposite signs we
immediately arrive at the combination (\ref{bg}).

We also point out that the third cumulant of the
current operator for noninteracting electrons
is known \cite{many} to be proportional to the parameters
(\ref{landr}) and (\ref{bg}) respectively at low and high voltages.
Following the same arguments as above we can {\it conjecture} that
the interaction
correction to the third cumulant should scale as $\beta -2\gamma$
at low voltages, while in the limit of large voltages one
can expect that this correction is governed by the
combination $\beta -6\gamma +6\delta$, where
\begin{equation}
\delta =\frac{\sum_nT_n^3(1-T_n)}{\sum_nT_n}.
\label{delta}
\end{equation}
This conjecture can also be generalized to higher cumulants of
the current operator.

We would like to emphasize that -- although the above conjecture
seems intuitively appealing -- it should still be verified by means of
a rigorous calculation which is beyond the scope of the present
paper. In the next section we will concentrate on the current noise
and will provide a detailed derivation of the results presented above.

\section{Effective action and current noise}

Similarly to Ref. \onlinecite{GZ00} we will use the effective action technique
in order to evaluate the current-current correlator for the system depicted in
Fig. 1. It is convenient to introduce the quantum phase variable $\varphi$
which is proportional to the integral of the fluctuating voltage (see Appendix
A). We will proceed within the Keldysh formalism and introduce two phase
variables $\varphi_{1,2}$ related to the two branches of the Keldysh contour.
Defining $\varphi^+=(\varphi_1+\varphi_2)/2$ and
$\varphi^-=\varphi_1-\varphi_2$ one can denote the overall phase jumps across
the scatterer as $\varphi^+ +eVt$ and $\varphi^-$. Correspondingly, the phase
jumps across the Ohmic shunt are $(eV_x-eV)t-\varphi^+$ and $-\varphi^-$. The
symmetric current-current correlation function (\ref{corr}) can
be expressed as follows
\begin{eqnarray}
\frac{1}{2}\langle \hat I (t)\hat I (t') + \hat I(t')\hat I (t)\rangle =-e^2
\int  {\cal D}\varphi^\pm \nonumber\\ \left[\frac{\delta^2}{\delta
\varphi^-_S(t)\delta\varphi^-_S(t')}+
\frac{\delta^2}{4\delta\varphi^+_S(t)\delta\varphi^+_S(t')}\right] e^{iS_{\rm
tot}[\varphi^\pm]}, \label{vder1}
\end{eqnarray}
see Appendix A for further discussion. By $\varphi^\pm_S$ we denote the phase
jumps over the ohmic shunt. The variational derivatives in Eq.(\ref{vder1})
act on the shunt part of action. Here $S_{\rm tot}[\varphi^{\pm}]$ is the
total action of our system
\begin{equation}
S_{\rm tot}[\varphi^{\pm}]=S[\varphi^{\pm}]+S_{S}[\varphi^{\pm}],
\end{equation}
where the term
\begin{eqnarray}
iS_S[\varphi^\pm]=\frac{i}{e^2 R_S} \int_0^\infty d t \varphi^-(t)
\left(\frac{eV_xR_S}{R+R_S}- \dot{\varphi}^+ (t)\right) \\ -\frac{1}{2e^2
R_S}\int_0^\infty dt_1 \int_0^\infty
dt_2\alpha(t_1-t_2)\varphi^-(t_1)\varphi^-(t_2)\nonumber
\end{eqnarray}
comes from the shunt, $S[\varphi^{\pm}]$ is the scatterer action
and
\begin{equation}
\alpha(t-t')=-\frac{1}{\pi}\left( \frac{\pi T}{\sinh[\pi T(t-t')]}\right)^2.
\label{alpha}
\end{equation}
A detailed derivation of the action $S[\varphi^{\pm}]$ is carried out
in Appendices A, B and C. The main idea of this derivation is
to expand $S[\varphi^{\pm}]$ in powers of $\varphi^-$ keeping
the full nonlinearity of the corresponding terms in $\varphi^+$.
This procedure is just the quasiclassical approximation for
the phase variable. It is parametrically justified under the
condition (\ref{metal}). In Ref. \onlinecite{GZ00} the action
$S[\varphi^{\pm}]$ was evaluated up to the second order in
$\varphi^{-}$. This is sufficient to derive the current-voltage
characteristics of the scatterer. However, in order to describe the
current noise it is necessary to expand the action $S[\varphi^{\pm}]$
further and to retain all terms up to the third order in $\varphi^-$
\begin{equation}
S[\varphi^{\pm}]=S^{(1)}+S^{(2)}+S^{(3)}.
\label{3}
\end{equation}
This expansion is analyzed in Appendix C. We will now use these
results and explicitly evaluate the current-current correlator
(\ref{corr}).

\begin{widetext}
\subsection{Contribution of first and second order terms}

Let us first restrict our attention to the contribution of
the first and second order terms in (\ref{3}). They read
\begin{eqnarray}
iS^{(1)}[\varphi^{\pm}]+iS^{(2)}[\varphi^{\pm}]=-\frac{i}{e^2} \int_0^\infty d
t\varphi^-(t)\left[C\ddot{\varphi}^+(t) + \frac{1}{R}\left(\dot{\varphi}^+(t)
+eV\right)  \right]\nonumber&&\\ -\frac{1}{2e^2 R}\int_0^\infty dt_1
\int_0^\infty dt_2\alpha(t_1-t_2)\varphi^-(t_1)\varphi^-(t_2)
\left\{1-\beta\nonumber +\beta\cos\left[
eV(t_1-t_2)+\varphi^+(t_1)-\varphi^+(t_2)\right]\right\}. &&\label{qact}
\end{eqnarray}
Employing
Eq. (\ref{vder1}) we obtain
\begin{equation}
\frac{1}{2}\langle \hat I(t)\hat I(t')+\hat I(t')\hat I(t)
 \rangle
=\frac{\alpha(t-t')}{R_S} +e^2 \langle K(t) K(t')\rangle,
\label{fn}
\end{equation}
where we defined
\begin{equation}
K(t)=\frac{1}{e^2 R_S}\left[eV_x-eV-\dot{\varphi}^+ +
i\int_0^\infty d \tilde t\alpha(t-\tilde t)\varphi^-(\tilde t) \right].
\label{fn1}
\end{equation}
Note that here we neglected terms which originate from the second variational
derivative with respect to $\varphi^+$ in (\ref{vder1}) since these terms are
smaller in the parameter $1/g_0^2$ than those kept. Angular brackets in Eq.
(\ref{fn}) imply averaging with the path integral
$$
\langle ... \rangle =\int{\cal D}\varphi^\pm (...)
\exp \left(iS^{(1)}[\varphi^\pm]+iS^{(2)}[\varphi^\pm]+iS_S[\varphi^\pm]\right).
$$
Rewriting the correlator $\langle KK\rangle$ as
\begin{equation}
\langle K(t) K(t')\rangle=-\lim_{\eta\rightarrow 0}\int{\cal D}\varphi^\pm
\frac{\delta^2}{\delta \eta(t)\delta\eta(t')} \exp\left\{
i\sum_{i=1,2}S^{(i)}[\varphi^\pm]+iS_S[\varphi^\pm]+
i\int_0^\infty d\tilde t \eta(\tilde
t)K(\tilde t)\right\}
\end{equation}
and performing a shift of $\varphi^-_\omega\to \varphi^-_\omega+
 \eta_\omega R_0/R_s(1-i\omega R_0C)$, we obtain
the expression for the Fourier transformed noise spectrum
(\ref{corr})
\begin{eqnarray}
{\cal S}_\omega&=&\omega\coth\frac{\omega}{2T} \left\{
\text{Re}\,\frac{1}{Z(\omega)}-\frac{\beta }{R\kappa\Omega}\right\}
+\frac{\beta
[\alpha(t-t')\langle\cos(eV(t-t')+\varphi^+(t)-
\varphi^+(t'))\rangle]_\omega
 }{R\kappa \Omega }+\label{appr} \\&& + \frac{\beta^2
}{e^2R^2\kappa \Omega}\left[\langle H(t)\rangle^2 -\langle
H(t)H(t')\rangle\right]_\omega ,\nonumber
\end{eqnarray}
where
\begin{equation}
H(t)=\int_0^\infty d\tilde t \alpha(t-\tilde t)\varphi^-(\tilde t)
\left[1-\cos\left(eV(t-\tilde t)+\varphi^+(t)-\varphi^+(\tilde t)\right)
\right]. \label{lf}
\end{equation}
In Eq. (\ref{appr}) we also introduced the following notations
\begin{equation}
Z(\omega)=R_S+\frac{1}{R^{-1}-i\omega C},\;\;\Omega =1+\omega^2R_0^2C^2,\;\;
\kappa=\frac{(R+R_S)^2}{R^2}.
\end{equation}

Making use of the relation (\ref{rel1}), we arrive at the correlator $\tilde S_\omega,$
\begin{equation}
\tilde {\cal S}_\omega=\frac{1}{R}\left((1-\beta)\omega\coth\frac{\omega}{2T}
+\beta [\alpha(t-t')\langle\cos(eV(t-t')+\varphi^+(t)-
\varphi^+(t'))\rangle]_\omega
\right)+\dots.
\label{tildeS}
\end{equation}
where $\dots$ stands for the terms containing $\left[\langle H(t)\rangle^2 -\langle
H(t)H(t')\rangle\right]_\omega$ in Eq. (\ref{appr}).

Within our approach interaction effects are described by the terms
containing the fluctuating variable $\varphi^+$. If one formally
sets this variable equal to zero, from (\ref{appr}) one immediately
recovers the noninteracting result  \cite{Khl,Les,Bu1}
\begin{equation}
\tilde {\cal S}_\omega^{\rm
ni}=(1-\beta)\frac{\omega}{R}\coth\frac{\omega}{2T}
+\frac{\beta}{2}\sum_{\pm}(\omega\pm
  eV)\coth\frac{\omega\pm eV}{2T}.
\label{Khlus}
\end{equation}
Taking the phase fluctuations into account we arrive at the expression
for the interaction correction to (\ref{Khlus}). However,
the corresponding expression turns out to be incomplete in two
respects. First, one of the terms does not satisfy
FDT. Second, the correction to (\ref{Khlus}) obtained in this way
scales with the parameter $\beta$ in both limits of small and
large voltages. While in the former limit (Nyquist noise) this result
is understandable and consistent with Ref. \onlinecite{GZ00}, at large
voltages (shot noise) one also expects an extra contribution.
Its existence can be anticipated because the shot noise is governed
by the combination $\sum_nT_n(1-T_n)$ and not simply by $\sum_nT_n$
as the Nyquist noise, see also our discussion in Sec. 2.

Both these problems are remedied by taking into account
the third order in $\varphi^-$ contribution to the effective action.
This will be demonstrated in the next subsection.

\subsection{Corrections due to third order terms}

Following the analysis in Appendix C we identify two different
contributions to the third order term
$$
S^{(3)}=S_{\beta}^{(3)}+S_{\gamma}^{(3)}.
$$
The first contribution has the form
\begin{equation}
iS^{(3)}_{\beta}[\varphi^\pm]=\frac{i\beta}{6e^2R}
\int\limits_0^\infty d\tau\; (\varphi^-(\tau))^3\dot\varphi^+(\tau).
\label{is31}
\end{equation}
Taking this term into account and repeating the above
analysis we arrive at an extra contribution to the current
noise in the form $-e^2\beta \delta(t-t')/2RC$, see also Appendix C.
Adding this contribution to Eq. (\ref{appr}) and subtracting
the noninteracting result (\ref{Khlus}) we arrive at the interaction
correction
\begin{equation}
\delta S_\omega^{(\beta )}=\frac{\beta}{R\kappa \Omega }
\left[\alpha(t-t')\cos\left(eV(t-t')\right)
\left( e^{-F(t-t') }-1 \right)-e^2\delta(t-t')/ 2C\right]_\omega .\label{final}
\end{equation}
The function $F(t)$ results from averaging over the phase
fluctuations
\begin{equation}
\langle\cos[eV(t-t')+\varphi^+(t)-
\varphi^+(t')]\rangle=\cos[eV(t-t')]e^{-F(t-t')}. \label{cav}
\end{equation}
This function has the form
\begin{equation}
F(t)=e^2R_0^2\int\limits_{-\infty}^{\infty}\frac{d\omega}{2\pi}\frac{1-
\cos\omega t}{\omega^2 \Omega }\left\{ \left( \frac{1}{R_0}-
\frac{\beta}{R}\right) \omega\coth\frac{\omega}{2T}
+\frac{\beta}{2R}\sum_{\pm} (\omega\pm eV)\coth \frac{\omega\pm eV}{2T}
\right\}. \label{fdt}
\end{equation}
We also note that in Eq. (\ref{final}) we omitted the last term of
Eq. (\ref{appr}) which contains averages of the function $H$
(\ref{lf}). Our analysis demonstrates that these terms are small in
all the regimes considered below.

What remains is to evaluate the correction to the shot noise
from the second contribution to $S^{(3)}$. The derivation of
this contribution is presented in Appendix C. Here we only
quote the result:
\begin{eqnarray}
iS^{(3)}_{\gamma}[\varphi^{\pm}]= \frac{\pi i\gamma T^3}{6e^2R}\int_0^ \infty
dy_1 \int_0^\infty dy_2\int_0^\infty dy_3\frac{\varphi^-(y_1)
\varphi^-(y_2)\varphi^-(y_3)}{\sinh[\pi Ty_{21}]\sinh[\pi Ty_{32}] \sinh[\pi
Ty_{13}]}\times\nonumber\\
\left\{\sin\left(eVy_{21}+\varphi^+(y_2)-\varphi^+(y_1)\right)+
\sin\left(eVy_{32}+\varphi^+(y_3)-\varphi^+(y_2)\right)+
\sin\left(eVy_{13}+\varphi^+(y_1)-\varphi^+(y_3) \right)
   \right\},\label{trp1}
\end{eqnarray}
where $y_{ij}=y_i-y_j$ and the parameter $\gamma$ is defined in
Eq. (\ref{gam}).

At the first glance this contribution to the effective action could be
considered unimportant. This is indeed the case in several
limits. For instance, at sufficiently small transmissions $\beta \gg
\gamma$ the term (\ref{trp1}) can obviously be neglected. In the limit
of low voltages one can, making use of the condition (\ref{metal}),
expand $S^{(3)}_\gamma$ in small phase fluctuations $\varphi^\pm$. Then one
gets $S^{(3)}_\gamma$ proportional to the combination
$(\varphi^+)^{3}(\varphi^-)^{3}$ which can be dropped as compared to
other terms provided $g_0 \gg 1$. However, in the limit of large
voltages the term (\ref{trp1}) gains importance and -- as we shall see
-- provides significant contribution to $\delta
{\cal S}_\omega$.

Proceeding along the lines with the above analysis we find
\begin{eqnarray}
& \delta {\cal S}^{(\gamma)}_\omega=&\frac{2\pi ^2 \gamma
T^3}{g_0R\kappa\Omega}\times \label{fa}\\ && \int_0^\infty d
t\int_0^\infty dx \frac{\left(1-e^{-x/R_0C}\right)\left( \cos [eVt] -\cos
[eVx]\right)\cos\omega t}{\sinh[\pi Tx]\sinh[\pi Tt]}\left(
\frac{1}{\sinh[\pi T(x-t)]}-\frac{1}{\sinh[\pi T(x+t)]}\right).\nonumber
\end{eqnarray}
This expression will be analyzed below in Sec. 3F.
\end{widetext}
\subsection{Relation to FDT}

Before we proceed with the analysis of the above results let us
establish some useful general expressions and illustrate the
relation between our approach and FDT. Throughout this subsection
we will only consider the limit of small voltages $eV\ll 1/R_0C$
and neglect the dependence of the function $F(t)$
on $V$. In the spirit of the $P(E)$-theory \cite{pe}
let us define the function
\begin{eqnarray}
P(E)=\int_{-\infty}^\infty dt e^{iEt}e^{-\Phi(t)},\\
\Phi(t)=F(t)|_{V=0}+\frac{ie^2R_0}{2}\text{sign}[t]\left(1-e^{-|t|/R_0C}
\right).\nonumber
\end{eqnarray}
This function obeys the ``detailed balance'' symmetry
$P(-E)=e^{-E/T}P(E)$ which follows from the property
$\Phi(t-(i/T))=\Phi(-t)$. Let us also introduce the function
\begin{eqnarray}
{\cal N}_\omega=\frac{1}{4\pi
R_0}\times\nonumber\\\sum_\pm\int_{-\infty}^{\infty}
dE\frac{E(1+e^{-(\omega\pm eV)/T})}{1-e^{-E/T}}P(\omega\pm eV-E), \label{io}
\end{eqnarray}
and rewrite it in the form
\begin{eqnarray}
{\cal N}_\omega=\frac{1}{R_0}\int_{-\infty}^{\infty}dt e^{i\omega t}\bigg\{
-\frac{e^2}{2 C}\delta(t)+ \nonumber\\\alpha(t)e^{-F(t)}\cos(eVt)\cos
\left[\frac{\pi}{g_0}\left(1-e^{-t/R_0C}\right) \right]  \bigg\}. \label{calS}
\end{eqnarray}
We observe that, since in the interesting for us limit $g_0 \gg 1$
the argument of
$\cos \left[\frac{\pi}{g_0}\left(1-e^{-t/R_0C}\right)\right]$ is
small, with the accuracy $\sim 1/g_0^2$ one can use the function
(\ref{calS}) in order to analyze the result (\ref{final}).

Proceeding further let us rewrite Eq. (\ref{io}) as
\begin{equation}
{\cal N}(\omega)=\frac{1}{2}\sum_{\pm}\coth\frac{\omega\pm eV}{2T}
{\cal I}(\omega\pm eV),
\label{sss}
\end{equation}
where
\begin{equation}
 {\cal I}(\omega)=\frac{1-e^{-\omega/T}}{2\pi R_0}\int_{-\infty}^\infty
\frac{dE E}{1-e^{-E/T}}P(\omega-E). \label{ti}
\end{equation}
After a simple algebra from Eq. (\ref{ti}) we obtain
${\cal I}(\omega)=\frac{\omega}{R_0} +\delta {\cal I} (\omega )$ and
\begin{eqnarray}
\delta {\cal I}(\omega)=\frac{2}{R_0}\times\nonumber\\\int_0^\infty dt
\sin(\omega t)e^{-F(t)} \alpha(t) \sin\left[\frac{\pi}{g_0}
\left(1-e^{-t/R_0C}\right) \right]. \label{tild}
\end{eqnarray}
Comparing the above expressions with Eq. (\ref{final})
we arrive at the following correction to the current noise
\begin{equation}
\delta {\cal S}_\omega =\frac{\beta R_0}{2R\kappa \Omega }
\sum_{\pm}\coth\frac{\omega\pm eV}{2T}\delta {\cal
I}(\omega\pm eV).
\end{equation}
In order to illustrate the relation between our results and FDT we
notice that in the relevant limit $g_0 \gg 1$
the quantity $\delta {\cal I} (eV)$ (\ref{tild}) is defined by exactly
the same time integral as the interaction correction to the $I-V$
curve, cf. Eq. (27) of Ref. \onlinecite{GZ00}. In particular, in the
limit of zero frequency and voltage one finds
 \begin{equation}
\delta \tilde {\cal S}_{\omega =0}=
2T e^2 \beta \int_0^\infty
t\alpha(t)e^{-F(t)}\left(1-e^{-t/R_0C}\right)d t.\label{ttt}
\end{equation}
In accordance with FDT the combination in the right hand side
is just the interaction correction to the zero-bias conductance of a
coherent scatterer \cite{GZ00}  multiplied by $2T$.

We will now derive the interaction correction to the current noise
in several important limits.

\subsection{High temperatures}
In the limit $T\gg 1/R_0C$ it is sufficient to evaluate the function
$F(t)$ only at short times $t\lesssim 1/T$. In this limit from Eq.
(\ref{fdt}) we get
\begin{equation}
F(t)=\frac{e^2t^2}{2C}\left[ \left(1 -\frac{\beta R_0}{R}
\right)T + \frac{\beta R_0}{2R}eV\coth\frac{eV}{2T} \right] .
\end{equation}
Expanding $e^{-F(t)}$ in Eq. (\ref{final}) to the first order in $F$ we obtain
\begin{eqnarray}
\delta {\cal S}_\omega =\frac{e^2\beta }{2RC\kappa \Omega}
\bigg\{-1+\nonumber\\ \left[1+\frac{\beta R_0}{R}\left(\frac{eV}{2T}
\coth\frac{eV}{2T}-1 \right)\right]\sum_{\pm}f\left(\frac{\omega\pm
eV}{2T}\right)\bigg\}, \label{ht}
\end{eqnarray}
where function $f(x)$ reads
\begin{equation}
f(x)=\frac{x\cosh x}{2\sinh^3 x}-\frac{1}{2\sinh^2 x}.
\end{equation}
In the limit of small frequencies and voltages we then find
\begin{equation}
\delta {\cal S}_\omega =-\frac{e^2 \beta }{3RC\kappa }. \label{offset1}
\end{equation}
At high frequencies $\omega\gg T, eV$ or large voltages $eV\gg
T,\omega$ we get
\begin{equation}
\delta {\cal S}_\omega =-\frac{e^2 \beta }{2RC\kappa \Omega}.
\label{offset2}
\end{equation}
Both results (\ref{offset1}) and (\ref{offset2}) describe
partial suppression of the current noise by Coulomb interaction.
As we have already discussed, Eq. (\ref{offset1}) is consistent
with the results \cite{GZ00} combined with FDT, whereas
Eq. (\ref{offset2}) just corresponds to the Coulomb offset $\Delta
V=-e\beta /2C$ on the $I-V$ curve of a coherent scatterer at large
voltages. For the sake of completeness we also note that in a
specific limit $|\omega\pm eV|\ll T$ Eq. (\ref{ht}) yields positive
correction to the current noise
\begin{equation}
\delta {\cal S}_\omega=\frac{e^2 \beta^2 R_0|eV|}{24CTR^2\kappa \Omega}.
\end{equation}
However, the magnitude of this
correction is small in the parameter $\sim e^2 R_0/(RCT)\ll 1/g_0$.

\subsection{Low temperatures}
Now let us consider the limit of low temperatures $T\ll 1/CR_0$.
At low voltages $eV \ll 1/R_0 C$ and times much longer than $1/R_0C$
the function $F(t)$ reads
\begin{equation}
F(t)\simeq \frac{2}{g_0}\ln\left( \frac{\sinh [\pi T t]}{\sinh [\pi T R_0 C]
}\right). \label{estf}
\end{equation}
Combining this expression with Eq. (\ref{final}), in the limit of
small $\omega ,eV< T$ we obtain
\begin{equation}
{\cal S}_\omega=2T\left[\frac{1}{R+R_S}-
\frac{\beta}{R\kappa}\left(1-(T R_0 C)^{\frac{2}{g_0}}\right)\right].
\label{coul1}
\end{equation}
For $T \gg g_0E_C\exp (-g_0/2)$ the result (\ref{coul1}) can be
expanded in $2/g_0$. In this limit for the interaction correction
we get
\begin{equation}
\delta {\cal S}_\omega=-\frac{4T\beta }{g_0R\kappa}\ln\frac{1}{R_0CT}.
\label{intt2}
\end{equation}
In the opposite limit of very low $T < g_0E_C\exp (-g_0/2)$ (but still
$T \gg \omega, eV$) the last
term in (\ref{coul1}) can be neglected and the interaction
correction becomes
\begin{equation}
\delta {\cal S}_\omega=-\frac{2T\beta }{R\kappa}.
\label{lowt2}
\end{equation}
In the limit $T\ll \omega,eV \ll 1/R_0 C$ we can
set $T=0$ in Eq. (\ref{estf}). Then we obtain
\begin{eqnarray}
{\cal S}_\omega=|\omega |\left( \frac{1}{R+R_S}- \frac{\beta
}{R\kappa}\right)+\nonumber\\\frac{\beta }{2R\kappa}\sum_\pm |\omega\pm
eV|\big[|\omega\pm eV|R_0C\big]^{\frac{2}{g_0}}. \label{coul2}
\end{eqnarray}
If both $\omega$ and $V$ tend to zero, the last term in (\ref{coul2})
can again be neglected and we find
\begin{equation}
\delta {\cal S}_\omega=-\frac{|\omega |\beta }{R\kappa}.
\label{lowo2}
\end{equation}
If, however, $\omega$ and/or $eV$ exceed the scale $g_0E_C\exp
(-g_0/2)$, one expands Eq. (\ref{coul2}) in $2/g_0$ and gets
\begin{equation}
\delta {\cal S}_\omega=-\frac{2|\omega |\beta }{g_0R\kappa}
\ln\frac{1}{|\omega|R_0C}.
\label{into2}
\end{equation}
This expression applies for $\omega \gg eV$. In the opposite limit
in Eq. (\ref{into2}) one should simply substitute $eV$ instead of
$\omega$. Note, however, that in the latter limit the corresponding
result yields only one contribution ($\delta S_\omega^{(\beta )}$)
to the interaction correction. Another contribution
($\delta S_\omega^{(\gamma )}$)
will be found in Sec. 3F.

To complete this subsection let us find the interaction
correction in the limit $\omega,eV\gg 1/R_0C$. At large voltages
the dependence of $F(t)$ on $V$ should be taken into account.
Evaluating the corresponding (linear in $V$) correction to $F(t)$
(\ref{estf}), we obtain
\begin{equation}
\delta {\cal S}_\omega=\frac{e^2 R_0^2\beta^2 |eV|}{4\pi R^2\kappa \Omega}
\sum_\pm w[(\omega\pm eV)R_0C],\label{lg}
\end{equation}
where
\begin{equation}
w(x)=-2-2x\arctan x -\ln x^2+\ln(1+x^2)+|x|\pi.
\end{equation}
For $|\omega\pm eV|\gg 1/R_0 C$ the asymptotics $w(x\gg 1)\simeq 1/3x^2$
should be used. In this case we again recover Eq. (\ref{offset2}). If,
however, $|\omega\pm eV|\alt 1/R_0C$, then
the interaction correction is governed by another asymptotics
$w(x\ll 1)\simeq -2-\ln x^2$ and, hence, this correction is
positive. Such an increase of the noise at
$|\omega\pm eV|<1/R_0C$ is similar to that found at higher temperatures.

\subsection{Large voltages}

Now let us evaluate the remaining correction $\delta {\cal S}^{(\gamma
  )}$ (\ref{fa}).
At high temperatures $T\gg 1/CR_0$ we obtain
\begin{eqnarray}
\delta {\cal S}^{(\gamma)}_\omega\sim \frac{e^2\gamma}{RC\kappa
\Omega}\left(\frac{eV}{T}\right)^2, \quad\text{if}\; \omega, eV\ll T ,\\
\delta {\cal S}^{(\gamma)}_\omega= \frac{e^2
\gamma}{RC\kappa\Omega}\theta\left(
|eV|-|\omega|\right)\tanh\frac{|eV|-|\omega|}{2T}, \\\text{if}\; \omega, eV\gg
T .\nonumber
\end{eqnarray}
In the limit  $\omega,eV\ll T\ll 1/CR_0$ one finds
\begin{equation}
\delta {\cal S}^{(\gamma)}_\omega= \frac{2\gamma (eV)^2}{3Tg_0R\kappa}\ln
\frac{1}{TR_0C}.
\end{equation}
Finally, at higher frequencies and voltages $\omega,eV\gg T$ we derive
\begin{eqnarray}
&\delta {\cal S}^{(\gamma)}_\omega=\frac{4\gamma}{g_0R\kappa\Omega}
\theta\left( |eV|-|\omega|\right)\bigg\{
\frac{\arctan[(|eV|-|\omega|)R_0C]}{R_0C}&\nonumber\\ &+
\frac{|eV|-|\omega|}{2}\ln \left[1+((|eV|-|\omega|)R_0C)^{-2}\right] \bigg\}
.& \label{ggg}
\end{eqnarray}
Note that the correction $\delta {\cal S}^{(\gamma)}_\omega$
is positive in all cases. As compared to previously obtained contribution
$\delta S^{(\beta )}_\omega$ the correction (\ref{ggg})
becomes important in the limit $eV\gg T,\omega$. For such voltages
both corrections add up, $\delta {\cal S}_\omega =
\delta {\cal S}^{(\beta )}_\omega +\delta {\cal S}^{(\gamma
  )}_\omega$, and yield
\begin{eqnarray}
\delta {\cal S}_\omega=-\frac{2(\beta-2\gamma)|eV|}{g_0R\kappa}\ln
\frac{1}{|eV|R_0C}, \label{summa}\\\nonumber \text{if}\; T,|\omega|\ll
|eV|\ll1/CR_0,
\\ \delta {\cal S}_\omega=-\frac{(\beta-2\gamma)
  E_C}{R\kappa\Omega},\quad\text{if}\; |eV|\gg T,1/CR_0,|\omega| . \nonumber
\end{eqnarray}
These results complete our analysis of current fluctuations
in coherent conductors with electron-electron interactions.

\section{Summary}

Combining the standard scattering matrix approach with the effective
action formalism we have analyzed the effect of electron-electron
interactions on current noise in mesoscopic coherent conductors
in the metallic limit (\ref{metal}).

We have found that Coulomb
interaction always leads to partial suppression of the
Nyquist noise. The corresponding interaction term is proportional
to the parameter $\beta$ (\ref{landr}) similarly to the interaction
correction to the conductance \cite{GZ00}. Interaction-induced
suppression of both conductance and Nyquist noise has the same
physical origin, and a direct relation between these two effects
can easily be established with the aid of FDT.

The effect of electron-electron interactions on the shot noise is
somewhat more complicated. In this case we have recovered two
different interaction corrections entering with opposite signs.
One of them is negative and it is again governed by the parameter
$\beta$. Another correction is positive and it is proportional to the
parameter $\gamma$ (\ref{gam}) which depends on the transmission
distribution in a different way. The net interaction correction
to the shot noise scales as $\delta {\cal S}_\omega (V) \propto
2\gamma -\beta$, i.e. it can be both negative and positive depending
on the relation between $\beta$ and $\gamma$. The contribution
to  $\delta {\cal S}_\omega (V)$ from the $n$-th conducting mode is
{\it negative} provided its transmission $T_n$ is smaller than
1/2 and it is {\it positive} otherwise. For coherent
diffusive conductors $2\gamma -\beta = -1/15$, i.e.
in this particular case electron-electron
interactions tend to decrease the shot noise.

The presence of two interaction corrections to the shot noise has a
transparent physical interpretation. The $\beta$-correction is
due to Coulomb blockade suppression of the Landauer conductance
(\ref{Landauer}) while the $\gamma$-correction originates from
the term $-\sum_n T_n^2$ in the expression for the shot noise
\cite{Khl,Les,Bu1}. The absolute value of this term is
also decreased by interactions. But, since it enters with the
negative sign, the corresponding contribution to the noise
spectrum turns out to be
positive. We believe that the effect of electron-electron
interactions on higher cumulants of the current operator can be
described in a similar manner.

This work is part of the Kompetenznetz `` Funktionelle Nanostructuren''
supported by the Landestiftung Baden-W\"urttemberg gGmbH.
One of us (A.V.G.) acknowledges support from the Alexander von Humboldt
Stiftung.

\begin{widetext}
\appendix
\section{Effective action and observables}
Following \cite{GZ00} let us combine the effective action formalism
\cite{SZ,GZ} with the usual Landauer scattering approach.
Within the latter approach one introduces a (relatively small)
scatterer which connects two bulk reservoirs. The scatterer is
described by the scattering matrix. In order to include
electron-electron interactions it is
necessary to reckon with the many-body Hamiltonian
\begin{equation}
\hat H=\int d\bm{r} \hat\Psi^+(\bm{r})\left[ -\frac{\nabla^2}{2m}+W(\bm{r})
\right]\hat\Psi(\bm{r})+\frac{1}{2}\int d\bm{r}\int
d\bm{r}'\hat\Psi^+(\bm{r})\hat\Psi^+(\bm{r}')
\frac{e^2}{|\bm{r}-\bm{r}'|}\Psi(\bm{r}')\Psi(\bm{r}).
\end{equation}
Here the term $W(\bm{r})$ accounts for boundary and impurity
potentials, external fields etc. After the standard
Hubbard-Stratonovich
decoupling of the interaction term one arrives at the following path
integral over an auxiliary field
$V(\bm{r},t)$
\begin{equation}
e^{-i\hat Ht}=\frac{\int{\cal D}V(\bm{r},t')\left[{\bf \hat T}\, e^{-i\int_0^t dt'
\hat H_{\rm eff}[V(\bm{r},t')]}\right]e^{i\int_0^tdt'\int d\bm{r} \frac{\nabla
V(\bm{r},t')^2}{8\pi}}}{\int{\cal D}V(\bm{r},t')e^{i\int_0^tdt'\int d\bm{r}
\frac{\nabla V(\bm{r},t')^2}{8\pi}}}. \label{hubstr}
\end{equation}
Here ${\bf \hat T}$ is the time-ordering operator and
\begin{equation}
\hat H_{eff}[V(\bm{r},t')]=\int d\bm{r} \hat\Psi^+(\bm{r})\left[
-\frac{\nabla^2}{2m}+W(\bm{r}) -eV(\bm{r},t)\right]\hat\Psi(\bm{r}).
\end{equation}
We choose to define the electron charge as $-e$.

The time dynamics of the density matrix $\rho$ is described by means
of the evolution operator $J$ defined on the Keldysh contour.
In what follows we shall denote the field $V$ on
the upper and lower parts of this contour by $V_{1,2}$.
The general expression for the density matrix reads
\begin{equation}
\rho(X_{1f},X_{2f},t_f)=\int dX_{1i}dX_{2i}
J\left(X_{1f},X_{2f};X_{1i},X_{2i};t_f,t_i\right)\rho(X_{1i},X_{2i},t_i),
\label{kern}
\end{equation}
where $X$ the set of relevant quantum degrees of freedom. We shall
assume that interaction with the fluctuating fields $V_{1,2}$ is
turned on at a time $t_i=0$. The time $t_f$ is supposed to be
large. Making use of (\ref{hubstr}) and integrating over the
fermionic degrees of freedom, we obtain
\begin{equation}
J=\int{\cal D}V_1 {\cal D}V_2\exp{iS[V_1,V_2]},
\end{equation}
where $S$ is the effective action
\begin{equation}
iS[V_1,V_2]=2\text{Tr}\,\ln\hat G_V^{-1}+i \frac{C}{2}\int_0^t dt'
[V_{LR1}^2-V_{LR2}^2],\label{effac}
\end{equation}
Here we defined $V_{LRj}=V_{Lj}-V_{Rj}$ and neglected the
spatial dependence of the fields $V_{L1,2}$ and $V_{R1,2}$
inside both the left (L) and the right (R) reservoirs. The
Green-Keldysh matrix $G_V(X_1,X_2)$ (here $X=(\bm{r},t)$) obeys the $2\times
2$ matrix equation
\begin{equation}
\left(i\frac{\partial}{\partial t_1}\hat{\mathbf{1}} -\hat{H}_0
(\mathbf{r}_1)\hat{\mathbf{1}} +e\hat V(X_1)\right) \hat G_V(X_1,X_2)=\delta
(X_1-X_2)\hat \sigma_z,
\label{keleq}
\end{equation}
where $\hat H_0=(-\nabla^2/2m)+W(\bm{r})$, $\hat V$ is a diagonal $2\times 2$
matrix with components $\hat V_{ij}=V_i\delta_{ij}$ and $\hat \sigma_z$ is the
Pauli matrix.

The above equation for the Green-Keldysh function should be
supplemented by the initial condition for the density matrix $\hat
\rho (t=0) =\hat \rho_0$, where $\hat \rho_0$ is the equilibrium
density matrix of noninteracting electrons. In what follows we will need
the solution of
Eq. (\ref{keleq}) in the absence of the fulcutating fields $V_{1,2},$ which reads
\begin{eqnarray}
G_{11}(t_1,t_2)&=&-i\theta(t_1-t_2)\hat U_1(t_1,t_2)+i\hat
U_1(t_1,0)\hat\rho_0\hat U_1(0,t_2), \nonumber\\
G_{22}(t_1,t_2)&=&-i\theta(t_2-t_1)\hat U_2(t_1,t_2)+i\hat
U_2(t_1,0)\hat\rho_0\hat U_2(0,t_2), \nonumber\\ G_{12}(t_1,t_2)&=&i\hat
U_1(t_1,0)\hat\rho_0\hat U_2(0,t_2), \nonumber\\ G_{21}(t_1,t_2)&=&-i\hat
U_2(t_1,0)[\hat 1-\hat\rho_0]\hat U_1(0,t_2), \label{Gij}
\end{eqnarray}
where $\hat U_{1,2}(t_1,t_2)$ are the evolution operators
\begin{equation}
\hat U_{1,2}(t_1,t_2)=\text{T}\,\exp\left[-i\int_{t_2}^{t_1} dt'\left(\hat H_0
-eV_{1,2}(\bm{r},t')\right)\right].
\end{equation}
One should keep in mind that in the operator products like $\hat
U\hat\rho\hat U$ integration over intermediate spatial coordinates is implied.

Instead of specifying $W(\bm{r})$ we will describe electron transfer
between the reservoirs by means of the scattering matrix formalism.
This procedure is standard and further details are provided in
Appendix B. In calculating the trace in
Eq.(\ref{effac}), we shall make an explicit integration over the longitudinal
coordinates. Integration over the transverse coordinates is replaced by
summing over the transmission channels of the scatterer. It is convenient
to introduce the phase variables
\begin{eqnarray}
\varphi^{+}(t)=\int_0^t dt'\left(eV_{LR1}(t')+eV_{LR2}(t')\right)/2,\\
\varphi^{-}(t)=\int_0^t dt'\left(eV_{LR1}(t')-eV_{LR2}(t')\right).
\end{eqnarray}
Provided the dimensionless conductance $g_0$ is large  $g \gg 1$
fluctuations of $\varphi^-(t)$ are strongly suppressed, so we can
expand the exact action $iS[\varphi^\pm]$ in powers of $\varphi^-$
keeping the full nonlinearity in $\varphi^+$. Note, that the
external voltage enters only in $\varphi^+$ but not in $\varphi^-$.
Hence, for the system of Fig. 1 we have to impose
the constraints $\sum_j\varphi^+_j(t)= eV_{x}t$,
$\sum_j\varphi^-_j(t)=0$.
Here and below in this Appendix the summation runs over different
elements in our circuit, i.e. the scatterer and the shunt.

Let us define the kernel of the current operator. Consider, e.g., the
upper part of the Keldysh contour. The
charge is proportional to the derivative of the action with the respect to the
external voltage. Due to the property $\delta S/\delta
V_1(\bm{r},t)=e\int_0^t\delta S/\delta\varphi_1(\bm{r},t)$ we can relate the
difference of the charges of the left and right reservoirs to the integral
of $\delta S/\delta\varphi_{LR}(t)$ over time. Hence, the latter
derivative is proportional to the current through the scatterer.
Keeping track of proper ordering in the upper part of
the Keldysh contour, for $t>t'$ we obtain
\begin{equation}
\langle \hat I_j(t)\hat I_j(t') \rangle=-e^2 \int {\cal
D}\varphi^\pm\frac{\delta^2}{\delta
\varphi_{1j}(t)\delta\varphi_{1j}(t')}e^{iS[\varphi^\pm]}.\label{upp}
\end{equation}
Combining (\ref{upp}) with a similar relation in the lower part of the
Keldysh contour we get
\begin{eqnarray}
\frac{1}{2}\langle \hat I_j (t)\hat I_j (t') + \hat I_j(t')\hat I_j
(t)\rangle=-\frac{e^2}{2}\int  {\cal D}\varphi^\pm
\left[\frac{\delta^2}{\delta \varphi_{1j}(t)\delta\varphi_{1j}(t')}+
\frac{\delta^2}{\delta \varphi_{2j}(t)\delta
\varphi_{2j}(t')}\right]e^{iS[\varphi^\pm]} .
\label{vder}
\end{eqnarray}
This equation is equivalent to (\ref{vder1}).

Below we shall proceed with an explicit calculation of the
action by defining the scattering states.

\section{Transmission channels}
Following the usual $\hat S$-matrix
approach \cite{Bue} let us introduce the transmission channels.
We will assume that far from the scatterer
the electron propagation in transverse and longitudinal directions can
be described separately. In this case the solution of the
Schr\"odinger equation
\begin{equation}
-\frac{\nabla^2}{2m}\psi(\bm{r})+W(\bm{r})\psi(\bm{r})=E\psi(\bm{r})
\end{equation}
 can be factorized
\begin{equation}
\psi(\bm{r})=\sum_n \; c_n \Phi_n(\bm{r}_\perp)\chi_n(x).
\end{equation}
Here $x$ is the coordinate along the lead and $\bm{r}_\perp$ are the
transverse coordinates. The transverse wave functions
$\Phi_n(\bm{r}_\perp)$ satisfy the equation
\begin{equation}
-\frac{\nabla^2_\perp}{2m}
\Phi_n(\bm{r}_\perp)+W(\bm{r}_\perp)\Phi_n(\bm{r}_\perp)=E_n\Phi(\bm{r}_\perp),
\end{equation}
where the subscript $n$ enumerates the
transmission channels (we are considering
only channels with $E_n<E_F$). The function $\chi_n(x)$ outside the scatterer
region is defined from the equation
\begin{equation}
-\frac{1}{2m}\frac{d^2}{dx^2}\chi_n(x)=(E-E_n)\chi_n(x).
\end{equation}

\begin{figure}
\includegraphics[width=4in]{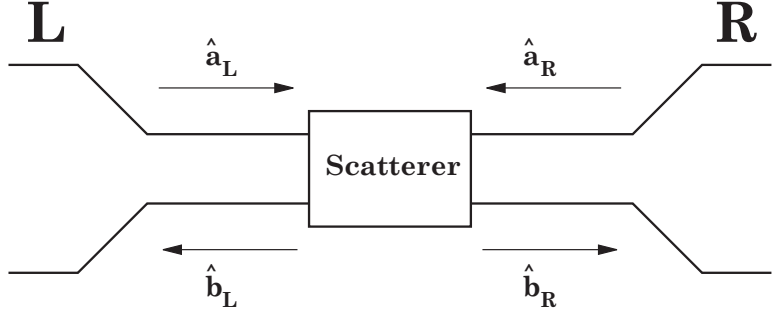}
\caption{Scattering states}
\end{figure}

Since the electronic states with energies $E$ close to the Fermi
energy $E_F$ mainly contribute, it is sufficient to describe the electron
dynamics quasiclassically. We define the energy $\xi=E-E_F$ and the
particle velocity in the $n$-th
channel $v_n=\sqrt{2(E_F-E_n)/m}.$ Then the wave function can be expressed as
\begin{eqnarray}
&\chi_n(x)=e^{i mv_nx}f^{in}_n(x)+e^{-imv_nx}f^{out}_n(x), & \text{left
reservoir.} \nonumber\\ &\chi_n(x)=e^{i
mv_nx}g^{out}_n(x)+e^{-imv_nx}g^{in}_n(x), & \text{right reservoir.}
\end{eqnarray}
In this way we have introduced the envelopes of the fast oscillating functions
$\exp(\pm imv_nx)$. Consider first the left
reservoir. The functions $f^{in}_n(x)$ and $f^{out}_n(x)$ satisfy the
following quasiclassical equations
\begin{eqnarray}
-iv_n\frac{d}{dx}f^{in}_n(x)=\xi f^{in}_n(x), \nonumber\\
iv_n\frac{d}{dx}f^{out}_n(x)=\xi f^{out}_n(x)
\end{eqnarray}
with the solutions
\begin{equation}
f^{in}_n(x)=\frac{e^{i\xi x/v_n}}{\sqrt{v_n}},\quad
f^{out}_n(x)=\frac{e^{-i\xi x/v_n}}{\sqrt{v_n}}.
\end{equation}
Analogously, for the right reservoir we find
\begin{equation}
g^{in}_m(x)=\frac{e^{-i\xi x/v_m}}{\sqrt{v_m}},\quad
g^{out}_n(x)=\frac{e^{i\xi x/v_m}}{\sqrt{v_m}}.
\end{equation}
The eigenfunction of the whole system with the energy $\xi$ in the left
reservoir may be expressed as
\begin{equation}
\psi_\xi(\bm{r}) = \sum\limits_n\left[ a_{Ln} e^{imv_nx} f^{in}_n(x) +b_{Ln}
e^{-imv_nx} f^{out}_n(x) \right]\Phi_n(\bm{r}_\perp), \label{psileft}
\end{equation}
while in the right reservoir we get
\begin{equation}
\psi_\xi(\bm{r}) = \sum\limits_k\left[ b_{Rk} e^{imv_kx} g^{out}_k(x) +a_{Rk}
e^{-imv_kx}g^{in}_k(x) \right]\Phi_k(\bm{r}_\perp). \label{psiright}
\end{equation}
The amplitudes of the outgoing $b_{L,R}$ and incoming $a_{L,R}$ states (see
the Fig. 2) are related via the scattering matrix $\hat S(\xi)$
\begin{equation}
\left( \begin{array}{c} b_{L1}\\ \cdots\\b_{LN_L}\\ b_{R1}\\ \cdots\\b_{RN_R}
\end{array}\right)=\hat S(\xi)
\left( \begin{array}{c} a_{L1}\\ \cdots\\a_{LN_L}\\ a_{R1}\\ \cdots\\a_{RN_R}
\end{array}\right).
\end{equation}
The unitary matrix $\hat S$ with dimensions $(N_L+N_R)\times(N_L+N_R)$ has the
block structure
\begin{equation}
\hat S(\xi) = \left( \begin{array}{cc} \hat r(\xi) & \hat t'(\xi)
\\ \hat t(\xi) & \hat r'(\xi)
\end{array}\right).
\label{Smatrix}
\end{equation}
The diagonal blocks $\hat r$ and $\hat r'$ describe reflection back to the
left and right reservoirs, respectively. The off-diagonal blocks describe
transmission through the scatterer. Later we shall neglect the
$\xi$-dependence of $\hat S$.

Let us now combine the incident $f^{in}_n(x)$ and outgoing $f^{out}_n(x)$ wave
functions belonging to the same channel into one wave function
$\psi_n(x)$. Namely, we assume that the scatterer is located at $x=0,$ and for
the left reservoir $(x<0)$ we put
\begin{equation}
\psi_n(y) = \left\{\begin{array}{ll} f^{in}_n(y), & y<0, \\ f^{out}_n(-y), &
y>0.
\end{array}\right.
\end{equation}
Analogously, for the right reservoir $(x>0)$ we define
\begin{equation}
\psi_m(y) = \left\{\begin{array}{ll} f^{in}_m(-y), & y<0, \\ f^{out}_m(y), &
y>0.
\end{array}\right.
\end{equation}
These new functions are defined in all the range $y\in [-\infty,+\infty]$ and
are equal to
\begin{equation}
\psi_j(y) = \frac{e^{i\xi y/v_j}}{\sqrt{v_j}}.
\end{equation}
Let us emphasize, that here the index $j$ enumerates all conducting
channels, both in the left and in
the right reservoirs (for convenience, we assume that the
left channels are enumerated first).

In the presence of the fluctuating field $V(t)$, the matrix elements of the
Hamiltonian in the reservoirs take the form
\begin{equation}
\hat H_{ij} = -i v_i\delta_{ij}\frac{\partial}{\partial y} - e
V_i(t)\delta_{ij},
\end{equation}
where $V_i=V_L$ for all left channels and $V_i=V_R$ for the right channels. If
at initial time $t_1$ the wave function was $\psi_n(t_1,y),$ then at the final
time $t_2>t_1$ it becomes
\begin{eqnarray}
\psi_n(t_2,y)&=&
e^{i[\varphi_n(t_2)-\varphi_n(t_1)]}\psi_n(t_1,y-v_n(t_2-t_1)), \;\;
y<0\;\;\text{or}\;\; y>v_n(t_2-t_1) ; \label{psiott}\\
\psi_n(t_2,y)&=&\sum\limits_k e^{i[\varphi_n(t_2)-\varphi_k(t_1)]
-i\left[\varphi_n\left(t_2-\frac{y}{v_n}\right)-\varphi_k\left(t_2-\frac{y}{v_n}
\right)\right]}s_{nk}\sqrt{\frac{v_k}{v_n}}\psi_k\left
(t_1,\frac{v_k}{v_n}y-v_k(t_2-t_1)\right),\: 0<y<v_n(t_2-t_1).\nonumber
\end{eqnarray}
Here $s_{nk}$ are the matrix elements of the $\hat S$-matrix and, as before,
$\varphi_n(t)=\int_0^t d\tilde t  eV_n(\tilde t)$. On the other hand, by
definition of the evolution operator we have
$$
\psi_n(t_2,y_2)=\sum\limits_k\int dy_1\; U_{nk}(t_2,t_1;
y_2,y_1)\psi_k(t_1,y_1).
$$
Comparing this expression with (\ref{psiott}), we find
\begin{eqnarray}
U_{nk}(t_2,t_1; y_2,y_1)&=&e^{i\varphi_n(t_2)} \bigg\{
\frac{\delta_{nk}}{v_n}\delta\left(\frac{y_2-y_1}{v_n}-t_2+t_1\right) +\;
\theta(y_2)\theta(v_n(t_2-t_1)-y_2) e^{-i\varphi_n\left(t_2-\frac{y_2}{v_n}
\right)} [s_{nk}-\delta_{nk}]\times \nonumber\\ && \sqrt{\frac{v_k}{v_n}}\;
e^{i\varphi_k\left(t_2-\frac{y_2}{v_n} \right)}
\delta\left(\frac{v_k}{v_n}y_2-y_1-v_k(t_2-t_1) \right)\bigg\}
e^{-i\varphi_k(t_1)}. \nonumber
\end{eqnarray}
It is convenient to introduce the new coordinates $\tau=y/v_n.$ More
precisely, instead of the wave function with the components $\psi_n(y)$ we
introduce the functions $\eta_n(\tau)=\sqrt{v_n}\psi_n(y/v_n).$ The kernels of
the operators will also be transformed. If the two functions are
related to each other by means of a linear operator
\begin{equation}
\psi_n^{(2)}(y)=\sum\limits_k \int dy'\; K_{nk}(y,y')\psi_k^{(1)}(y'),
\end{equation}
then the corresponding wave functions $\eta^{(2)}$ and $\eta^{(1)}$
satisfy the following relation:
\begin{equation}
\eta^{(2)}_n(\tau)=\sum\limits_k\int d\tau'\; \tilde
K_{nk}(\tau,\tau')\eta^{(1)}_k(\tau'),
\end{equation}
where
\begin{equation}
\tilde K_{nk}(\tau,\tau')=\sqrt{v_n}K_{nk}(v_n\tau,v_k\tau')\sqrt{v_k}.
\label{Ktr}
\end{equation}
In this representation the evolution operator can be simplified. We find
\begin{eqnarray}
\hat U(t_2,t_1;\tau_2,\tau_1)=\delta(\tau_2-\tau_1-t_2+t_1)
e^{i\hat\varphi(t_2)}\bigg\{ \hat 1+ \theta(\tau_2)\theta(-\tau_1)
e^{-i\hat\varphi(t_2-\tau_2)}[\hat S-\hat 1] e^{i\hat\varphi(t_1-\tau_1)}
\bigg\} e^{-i\hat\varphi(t_1)}. \label{U}
\end{eqnarray}
The matrix $\hat\varphi$ is diagonal with respect to the channel indices
$\hat\varphi_{ik}=\varphi_i\delta_{ik}$.  We also obtain an expression for the
inverse operator, i.e. the operator defined by $\int d\tau_2\hat
U(t_2t_1;\tau_3\tau_2) \hat
U^{-1}(t_2t_1;\tau_2\tau_1)=\delta(\tau_3-\tau_1)$.
It reads
\begin{eqnarray}
\hat U^{-1}(t_2,t_1;\tau_2,\tau_1)=\delta(\tau_1-\tau_2-t_2+t_1)
e^{i\hat\varphi(t_1)}\bigg\{ \hat 1+ \theta(\tau_1)\theta(-\tau_2)
e^{-i\hat\varphi(t_1-\tau_2)}[\hat S^+-\hat 1] e^{i\hat\varphi(t_2-\tau_1)}
\bigg\} e^{-i\hat\varphi(t_2)}. \label{U+}
\end{eqnarray}
Eqs. (\ref{U}), (\ref{U+}) apply for $t_2>t_1$, in order
to construct the corresponding expressions in the
opposite case one should just use the property
$\hat U (t_2,t_1;\tau_2,\tau_1)=\hat U^{-1}(t_1,t_2;\tau_2,\tau_1)$.

Finally let us define the equilibrium density matrix for
noninteracting electrons. It can be written in the form
\begin{equation}
\rho_{0,nk}(y_1,y_2)=\delta_{nk}\int\frac{dp}{2\pi}\;\frac{
e^{ip(y_1-y_2)}}{1+ e^{pv_n/T}}=\frac{\delta_{nk}}{2}\delta(y_1-y_2)
-\frac{\delta_{nk}}{2\pi} \frac{\pi i T}{v_n\sinh\left[ \frac{\pi
T(y_1-y_2)}{v_n}\right]}
\end{equation}
Performing the transformation (\ref{Ktr}) we obtain
\begin{equation}
\hat\rho_{0}(\tau_1,\tau_2)=\frac{1}{2}\left(\delta (\tau_1-\tau_2 )- \frac{i
T}{\sinh\left[ \pi T(\tau_1-\tau_2)\right]}\right) \hat 1. \label{rho0}
\end{equation}

\section{Expansion in the phase difference}
We shall expand the effective action (\ref{effac}) perturbatively in
$\varphi^-$. The field $\varphi^+$ will be taken into account exactly in each
term of this expansion. The expansion starts from the first order in
$\varphi^-$, since for $\varphi^-=0$ the contributions
from the forward and backward parts of the Keldysh contour
cancel each other. We get from Eq. (\ref{keleq})
\begin{equation}
2\text{Tr}\,\ln\hat G_V^{-1}=2\text{Tr}\ln\left( 1+\frac{\hat
G_0\hat{\dot{\varphi}}^-}{2}\right).\label{expan}
\end{equation}
The Green-Keldysh matrix $\hat G_0$ is evaluated for
$\hat\varphi^-=0$, i.e. it
is defined by Eqs. (\ref{Gij}) with the evolution operator
(\ref{U}) taken at $\hat \varphi =\hat\varphi^+$).
The fluctuating field $\hat
\varphi^-$ in  (\ref{expan}) is a unity matrix in Keldysh space and
a diagonal matrix in the space of conducting channels. Performing the
summation over the Keldysh indices we arrive at the first order in $\varphi^-$
contribution to the action $iS^{(1)}$
\begin{equation}
iS^{(1)}[\varphi^\pm]=-i\int dt\int ds\int d\tau_1\int
d\tau_2\,\text{Tr}\,\left[ \hat U(t,0;s,\tau_1)\left\{ \frac{1}{\pi}\frac{\pi
i T} {\sinh[\pi T(\tau_1-\tau_2)]} \right\} \hat
U^{-1}(t,0;\tau_2,s)\hat{\dot\varphi}^-(t)\right].
\label{corr1}
\end{equation}
For simplicity in Sec. 3 we have set  $t_f\to\infty$.
Here we will keep it finite and use the
conditions $\varphi^-(0)=\varphi^-(t_f)=0$. The $\delta$-functions
contained in $\hat U$-matrices of (\ref{corr1}) will lead to a
singularity of the form $1/\sinh[\pi
T(\tau_1-\tau_2)]$ which is cured as follows. Let us choose close but
not exactly equal arguments $s_1$ and $s_2$. Expanding the combination
\begin{equation}
-i\int dt\int d\tau_1\int d\tau_2\, \text{Tr}\,\left[ \hat U(t,0;s_1,\tau_1)
\hat U^{-1}(t,0;\tau_2,s_2)\hat{\dot\varphi}^-(t)\right]
\end{equation}
to the first order in $s_1-s_2$ and multiplying the result by
$1/\sinh[\pi T(s_1-s_2)]$, we obtain
\begin{equation}
iS^{(1)}[\varphi^\pm]=\frac{i}{\pi}\int_0^{t_f} ds \text{Tr}\,\left[
\hat\varphi^-(s)\left(\hat S\hat{\dot{\varphi}}^+(s)\hat S^+-
\hat{\dot{\varphi}}^+(s)\right)\right].
\end{equation}
Making use of the condition $\text{Tr}\,[\hat{t}^+\hat
  t]=\text{Tr}\,[\hat{t'}^+\hat t']$ we get
\begin{equation}
iS^{(1)}[\varphi^\pm]=-\frac{i}{\pi}\text{Tr}\,[\hat t^+\hat
t]\int\limits_0^{t_f} d\tau\; \varphi^-(\tau)\dot\varphi^+(\tau),
\label{SRRR}
\end{equation}
where $\varphi^\pm(\tau) = \varphi^\pm_L(\tau)-\varphi^\pm_R(\tau)$.

Consider now the contribution to the action of the
second order in $\varphi^-$. It is defined as
\begin{equation}
iS^{(2)}[\varphi^\pm]=-\text{Tr}\,\left[\hat G_{12}\hat{\dot\varphi}^-\hat
G_{21}\hat{\dot\varphi}^- \right].
\end{equation}
After a straightforward algebra one obtains
\begin{eqnarray}
iS^{(2)}[\varphi^\pm]&=-
&\int\limits_{0}^{t_f}d\tau_1\int\limits_{0}^{t_f}d\tau_2
\rho_0(\tau_2-\tau_1)\rho_0^*(\tau_1-\tau_2)\times \nonumber\\ && \text{ Tr}\,
\bigg\{ e^{-i[\hat\varphi^+(\tau_1)-\hat\varphi^+(\tau_2)]} \big[\hat
S^+\hat\varphi^-(\tau_1)\hat S-\hat\varphi^-(\tau_1)\big]
e^{i[\hat\varphi^+(\tau_1)-\hat\varphi^+(\tau_2)]} \big[\hat
S^+\hat\varphi^-(\tau_2)\hat S-\hat\varphi^-(\tau_2)\big]\bigg\}.
\end{eqnarray}
Taking  into account the block structure of the $\hat S$-matrix, we find
\begin{eqnarray}
iS^{(2)}[\varphi^\pm]&=&-\frac{1}{4\pi}\int\limits_{0}^{t_f}d\tau_1
\int\limits_{0}^{t_f}d\tau_2\alpha(\tau_1-\tau_2)\bigg\{
\left[\text{Tr}\,(\hat{t'}^+\hat t')^2 + \text{Tr}\,(\hat{t}^+\hat t)^2
\right]\varphi^-(\tau_1)\varphi^-(\tau_2)+\nonumber\\&& 2\text{Tr}\,\left[
\hat r'\hat{r'}^+\hat t \hat t^+ \right]\cos \left(
\varphi^+(\tau_1)-\varphi^+(\tau_2)\right) \varphi^-(\tau_1)\varphi^-(\tau_2)
\bigg\}, \label{is2}
\end{eqnarray}
where $\alpha(\tau)$ is defined in Eq. (\ref{alpha}). Introducing the
parameter $\beta =\text{Tr}\,\left[
\hat r'\hat{r'}^+\hat t \hat t^+ \right]/\text{Tr}\,\left[
\hat t \hat t^+ \right]$ we rewrite Eq. (\ref{is2}) in a more
compact form
\begin{equation}
iS^{(2)}[\varphi^\pm]=-\frac{\text{Tr}\,\left[\hat{t}^+\hat
t\right]}{2\pi}\int\limits_{0}^{t_f}d\tau_1
\int\limits_{0}^{t_f}d\tau_2\alpha(\tau_1-\tau_2)
\varphi^-(\tau_1)\varphi^-(\tau_2)\left[1-\beta+\beta \cos \left(
\varphi^+(\tau_1)-\varphi^+(\tau_2)\right) \right].
\label{comp}
\end{equation}

We now proceed to the third order contribution $iS^{(3)}$ to the effective
action. It reads
\begin{equation}
iS^{(3)}=\frac{i}{2}\text{Tr}\,\left[\int\limits_{0}^{t_f}dt_{1}
\int\limits_{0}^{t_f} dt_2 \int\limits_{0}^{t_f} dt_3\theta(t_1-t_2)
\theta(t_3-t_2)\hat{\cal F}(t_1)\hat{\cal F}(t_2) \hat{\cal
F}(t_3)\hat\rho_0^{as} \right]+\frac{1}{12}\text{Tr}\,\left[\left(\hat
G_{12}\hat{\dot{\varphi}}^-+ \hat G_{21}\hat{\dot{\varphi}}^-\right)^3\right].
\label{s3f}
\end{equation}
Here we used the notations
\begin{equation}
\hat\rho_0^{as}(\tau_1-\tau_2)=- \frac{i T}{2\sinh\left[ \pi
T(\tau_1-\tau_2)\right]} \hat 1,\quad \hat{\cal F}(t)=\hat
U^{-1}(t,0)\hat{\dot{\varphi}}^-(t)\hat U(t,0).
\end{equation}
We obtain two terms from Eq. (\ref{s3f}). The first one is
\begin{equation}
iS^{(3)}_\beta [\varphi^\pm]=\frac{i\beta}{6\pi}\text{Tr}\,[\hat t^+\hat
t]\int\limits_0^{t_f} d\tau\; (\varphi^-(\tau))^3\dot\varphi^+(\tau).
\label{is3}
\end{equation}
In deriving this result we employed the same -- although somewhat more
involved -- regularization procedure
as for the first order contribution $S^{(1)}$. This procedure allows to
determine the correct overall prefactor in (\ref{is3}). One can then
verify that the resulting effective action satisfies the requirements
of FDT.

The second term, coming from Eq. (\ref{s3f}),
has the form
\begin{eqnarray}
iS^{(3)}_\gamma [\varphi^\pm]=\frac{4}{3}\text{Tr}\,\left[ \left(\hat t\hat
t^+\right)^2\hat r'\hat{r'}^+ \right]\int_0^{t_f} dy_1 \int_0^{t_f}
dy_2\int_0^{t_f}
dy_3\rho_0^{as}(y_2-y_1)\rho_0^{as}(y_3-y_2)\rho_0^{as}(y_1-y_3)
\times\nonumber\\ \varphi^-(y_1) \varphi^-(y_2)\varphi^-(y_3)
\left\{\sin\left(\varphi^+(y_2)-\varphi^+(y_1)\right)+
\sin\left(\varphi^+(y_3)-\varphi^+(y_2)\right)+
\sin\left(\varphi^+(y_1)-\varphi^+(y_3) \right)
   \right\}.\label{trp}
\end{eqnarray}
Defining the parameter $\gamma =\text{Tr}\,\left[
\left(\hat t\hat t^+\right)^2\hat r'\hat{r'}^+\right]/\text{Tr}\,\left[
\hat t \hat t^+ \right]$ and shifting the phase $\varphi^+$ by $eV$ we
obtain Eq. (\ref{trp1}). Collecting now all four contributions
(\ref{SRRR}), (\ref{comp}), (\ref{is3}) and (\ref{trp}) we arrive
at the final result for the effective action
\begin{equation}
S=S^{(1)}+S^{(2)}+S^{(3)}_\beta +S^{(3)}_\gamma .
\label{everything}
\end{equation}
This action is valid up to the third order in
$\varphi^-$, and the variable $\varphi^+$ is treated exactly
in each of the terms in (\ref{everything}).

It is instructive to compare our results with the AES action
\cite{AES} derived for tunnel junctions ($\beta\to 1$) to all orders in
$\varphi^\pm$. Rewriting the action \cite{AES} in our notations
together with the capacitive term one has
\begin{eqnarray}
iS_{AES}&=&\frac{4i}{ e^2 R} \int_0^{t_f} dt_1\int_0^{t_f}
dt_2\alpha_I(t_1-t_2) \theta(t_1-t_2)
\sin(\varphi^+(t_1)-\varphi^+(t_2))\sin\frac{
\varphi^-(t_1)}{2}\cos\frac{\varphi^-(t_2)}{2} \label{aes}\\&&-\frac{2}{e^2
R}\int_0^{t_f} dt_1\int_0^{t_f} dt_2\alpha(t_1-t_2) \sin\frac{
\varphi^-(t_1)}{2}\sin\frac{\varphi^-(t_2)}{2}
\cos(\varphi^+(t_1)-\varphi^+(t_2))-\frac{i}{e^2} \int_0^{t_f} d t
C\ddot{\varphi}^+\varphi^-.
  \nonumber
\end{eqnarray}
Here we denoted $\alpha_I(t_1-t_2)=\delta'(t_1-t_2)$. This
$\delta$-function should be understood as a smeared one.

Let us expand (\ref{aes}) in $\varphi^-$ and compare with our results
order by order. The first order terms are exactly the same for both
models. The difference between the models shows up in the second order
terms, for our model the parameter $\beta$ appears
explicitly in the second order contribution (\ref{comp}). In the limit $\beta
\to 1$ this expression reduces to that obtained from (\ref{aes}).
Expanding the action (\ref{aes}) to the third order in $\varphi^-$ one
only recovers the term of the form (\ref{is3}) with $\beta =1$, while
another term (\ref{trp}) cannot be recovered. Contributions of this
nature are not contained in the AES action at all since they are
proportional to higher orders of the channel transmission $T_n$.

It is worthwhile to point out that a formally exact representation
for the effective action
of a coherent scatterer (all orders in $T_n$ and all orders in
$\varphi^{\pm}$) can also be derived \cite{SZ,Z,Naz,GZ02}.
However, this formal expression turns out to be quite complicated to
deal with in the situation addressed here. For $g=R_q/R \gg 1$ and
provided instanton effects \cite{Naz,GZ02} can be neglected all
necessary information is equally contained in a much simpler form
of the effective action derived in the present paper.

We also note that there exists a simple relation between
the action derived here and the cumulant generating function
describing the full counting statistics of the charge transport in
noninteracting coherent conductors \cite{LLL}. This relation can be
established if one neglects fluctuations of the phase variable,
i.e. sets  $\varphi^+=eVt,$ and chooses $\varphi^-$ to be time
independent. By identifying $\varphi^-=-\lambda$ and expanding
the generating function $\ln(\chi(\lambda))$ (defined in Eq. (37)
of Ref. \onlinecite{LLL}) one arrives at the following indentity:
$$
\ln(\chi(\lambda))=\frac{i}{2}S[eVt,-\lambda].$$

Finally, let us use the AES action (\ref{aes}) in order to illustrate
the importance of the third order in $\varphi^-$ terms for the
calculation of the currect-current correlation functions.
Applying Eq. (\ref{vder}) one gets a contribution
stemming from the double differentiation of the term with $\alpha_I$ in the
action (\ref{aes})
\begin{equation}
\delta {\cal S}(t,t')=-\left\langle\frac{i}{R}\theta(t-t')\alpha_I(t-t')
\sin(\varphi^+(t)-\varphi^+(t'))\sin\frac{\varphi^-(t)-
\varphi^-(t')}{2}\right\rangle+(t\leftrightarrow t'). \label{noiadd}
\end{equation}
Evaluating this average with the aid of path integrals one has to keep
all nonlinear terms in the pre-exponent. However, the dependence of
the cosine term on $\varphi^+$ in the action in the exponent
can be neglected provided $g \gg
1$. Applying the identity
$\sin(\varphi^+(t)-\varphi^+(t'))=\sum_{\nu=\pm}e^{i\nu[\varphi^+(t)-
\varphi^+(t')]} /2i$ we arrive at the following integral
\begin{equation}
\int{\cal D}\varphi^+\exp\left\{i\nu
(\varphi^+(t)-\varphi^+(t'))-\frac{i}{e^2} \int_0^\infty d\tilde
t\varphi^+(\tilde t)\left[ C\ddot{\varphi}^-(\tilde t)
-\frac{\dot{\varphi}^-(\tilde t)}{R}\right] \right\},\label{fudel}
\end{equation}
which yields the $\delta$-function
\begin{equation}
\varphi^-(\tilde t)=\frac{2\pi}{g}\nu\left[\theta(t-\tilde t)\left(1-e^{-(t-\tilde
t)/RC}\right) - \theta(t'-\tilde t)\left(1-e^{-(t'-\tilde
t)/RC}\right)\right].\label{rel}
\end{equation}
\end{widetext}
As it was expected, $\varphi^-(\tilde t)$ is indeed small for $g \gg 1$.
Combining the above expressions we arrive at the contribution
$$
\delta {\cal S}(t,t')=-e^2\delta(t-t')/2RC.
$$
The same contribution multiplied by the factor $\beta$ was derived in
Sec. 3B from the term $S^{(3)}_\beta$ (\ref{is31}).
The above analysis makes the significance of the third order
in $\varphi^-$ terms in the action particularly transparent: The
kernel $\alpha_I(t)$ introduces the time-derivative of $e^{-t/RC}$
which compensates for an extra smallness $\sim 1/g$.


\begin{thebibliography}{}
\bibitem{FN} In this paper we neglect instanton effects \cite{Naz,GZ02}.
These effects are fully suppressed for $R_S <R_q$, while in the opposite
case $R_S > R_q$ they may become significant only at exponentially
small energies.
\end{thebibliography}
\end{document}